\documentclass[aps,twocolumn,showpacs]{revtex4}
\usepackage{amssymb}
\usepackage{amsfonts}
\usepackage{amsmath}
\usepackage{mathrsfs}
\begin{document}

\title{Tensor gauge condition and tensor field decomposition}

\author{Xiang-Song Chen$^{1,2,3,}$}
\email{cxs@hust.edu.cn}
\author{Ben-Chao Zhu$^1$}

\affiliation{$^1$Department of Physics, Huazhong
University of Science and Technology, Wuhan 430074, China\\
$^2$Joint Center for Particle, Nuclear Physics and
Cosmology,
Nanjing 210093, China\\
$^3$Kavli Institute for Theoretical Physics China, Chinese
Academy of Science, Beijing 100190, China}

\date{\today}

\begin{abstract}
We discuss various proposals of separating a tensor field
into pure-gauge and gauge-invariant components. Such tensor
field decomposition is intimately related to the effort of
identifying the real gravitational degrees of freedom out
of the metric tensor in Einstein's general relativity. We
show that, as for a vector field, the tensor field
decomposition has exact correspondence to, and can be
derived from, the gauge-fixing approach. The complication
for the tensor field, however, is that there are infinitely
many complete gauge conditions, in contrast to the
uniqueness of Coulomb gauge for a vector field. The cause
of such complication, as we reveal, is the emergence of a
peculiar gauge-invariant pure-gauge construction for any
gauge field of spin $\ge 2$. We make an extensive
exploration of the complete tensor gauge conditions and
their corresponding tensor field decompositions, regarding
mathematical structures, equations of motion for the
fields, and nonlinear properties. Apparently, no single
choice is superior in all aspects, due to an awkward fact
that no gauge-fixing can reduce a tensor field to be purely
dynamical (i.e., transverse and traceless), as can the
Coulomb gauge in a vector case.

\end{abstract}

%11.15.-q Gauge field theories
%04.20.Cv Fundamental problems and general formalism
\maketitle

\section{Introduction}

It is a familiar practice to separate a vector field $\vec
A$ into transverse and longitudinal parts, $\vec A\equiv
\vec A_\perp +\vec A_{\parallel}$, defined by $\vec
\partial \cdot \vec A_\perp=0$ and $\vec
\partial \times \vec A_\parallel=0 $. One may naturally think of
an analogous separation for a tensor field. Such field
decompositions are well motivated in physics. For the
vector case, e.g., it is the transverse field $\vec
A_\perp$ that describes a real photon. The need of
decomposing the tensor field is closely related to gravity.
The Einstein equivalence principle dictates that
gravitational effect is described by the metric tensor,
which also characterizes the inertial effect associated
with coordinate choice. But sometimes, one does face the
necessity of identifying the real gravitational degrees of
freedom out of the metric, e.g., in associating a
meaningful energy to gravitational radiation, in analyzing
canonical structure of gravitation and thus quantizing it
to define a physical graviton. The attempt to separate
gravity from inertial effect dates back to the bimetric
theory of Rosen in 1940 \cite{Rose40}. This theory is yet
{\em formal} because Rosen does not give any actual
prescription to separate a background metric from the
experimentally measured total metric. After about 20 years,
Arnowitt, Deser and Misner (ADM) proposed the famous
transverse-traceless (TT) decomposition of a symmetric
tensor in their canonical formulation of general relativity
\cite{ADM}. Based upon the quantum action principle,
Schwinger presently a slightly different TT decomposition
in his attempt of quantizing the gravitational field
\cite{Schw63}. Later on, such tensor decomposition was
further developed by Deser \cite{Dese67}, and York
\cite{York73}, with the aim of going covariantly beyond the
linear approximation. These decompositions are really {\em
operational} by giving explicit expression of the separated
component in terms of the total tensor.

Separating the metric tensor is much more involved than
separating $\vec A$ into $\vec A_\perp +\vec
A_{\parallel}$, due to one more index and also the
nonlinearity of gravity. Regarding nonlinearity, it should
be mentioned that in the Yang-Mills theory separating the
physical degrees of freedom from the gauge freedom is not
so trivial either. This problem has revived recently in
connection with the nucleon spin structure
\cite{SpinReview}. A possible solution is presented by Chen
{\it et al.} \cite{Chen08,Chen09}, and is extended
successfully to gravity by Chen and Zhu (CZ)
\cite{Chen11d}.

The decompositions of ADM, Schwinger, Deser, York, and CZ
are motivated from different aspects, and thus naturally
contain notable differences. The CZ decomposition is
specifically aimed at a clear separation of a pure-gauge
background, while the others are closely associated with
dynamical structure of gravity. In this paper, we explore
and compare these decompositions from several perspectives.
After a brief review of these decompositions, we first look
at the relatively simple linearized gravity, for which we
show that the aforementioned decompositions can all be
conveniently rederived from a gauge-fixing approach. In so
doing, we reveal an interesting mathematical structure that
the decompositions of ADM, Schwinger, Deser and York
contain doubly-nonlocal operation (namely, with the inverse
Laplacian operator $\frac 1 {\vec
\partial ^2}$ used quadratically), while that of CZ is the
unique singly-nonlocal choice. We then go beyond the linear
order and demonstrate that the CZ decomposition differs
critically from those of Deser and York in their physical
implications: The physical and pure-gauge (or inertial)
effects are cleanly kept apart to all orders in the CZ
formulation, but start to mix in the Deser and York
approaches beyond linear order. We also explain how the
Deser decomposition can be modified in the spirit of CZ
formulation. Finally, we give some further discussion about
how to analyze the physical and dynamical content of
gravity, regarding the selection of constraints (or
gauge-fixing) and the equations of motion. Apparently,
there is no universally superior choice of gauge for a
tensor field. The significant complication from the vector
case to the tensor case (and virtually all higher-rank
tensors) is revealed to arise from two facts: (i) It is
possible to construct a peculiar gauge-invariant pure-gauge
field, thus there are infinitely many ways to fix the gauge
completely; (ii) however, no gauge-fixing can pick out
directly the purely dynamical (i.e., transverse and
traceless) component of a tensor, and extra extraction is
needed.

\section{Various tensor decompositions}
The original TT decomposition of ADM is linear and for a
symmetric {\em spatial} tensor:
\begin{equation}
h_{ij} \equiv h_{ij}^{TT}+h_{ij}^T+h_{ij}^{L}, \label{ADM}
\end{equation}
where $h_{ij}^{TT}$ is TT: $h_{ij,i}^{TT}=h_{kk}^{TT}=0$,
$h_{ij}^T\equiv\frac 12 (\delta _{ij} h^T -\frac 1{\vec
\partial ^2} h^T_{,ij})$ is constructed to be transverse:
$h^T_{ij,i}=0$, and $h_{ij}^L\equiv f_{i,j}+ f_{j,i}$ is
longitudinal (pure-gauge). (Notations: Greek indices run
from 0 to 3, Latin indices run from 1 to 3, and repeated
indices are summed over, even when they both appear raised
or lowered. $\partial$ or comma denotes ordinary
derivative, and $\nabla$ denotes covariant derivative.) The
four unknowns, $h^T$ and $f_i$, are solved as follows:
$h_{ij,j}^{L}=h_{ij,j}$ gives $f_i$, then
$h_{kk}^T+h_{kk}^{L}=h_{kk}$ gives $h^T$. The results are:
\begin{eqnarray}
f_i&=& \frac 1{\vec \partial ^2}
(h_{ik,k} -\frac 12 \frac 1{\vec \partial ^2} h_{kl,kli}), \label{fi}\\
h^T&=&h_{kk}-\frac 1{\vec \partial ^2} h_{kl,kl}.
\end{eqnarray}

Inserting them into Eq. (\ref{ADM}) gives the explicit form
of the TT component:
\begin{eqnarray}
&&h_{ij}^{TT}=h_{ij}-\frac 12 \delta _{ij}(h_{kk}-\frac
1{\vec \partial ^2} h_{kl,kl}) \nonumber \\&&-\frac 1{\vec
\partial ^2}(h_{ik,kj}+h_{jk,ki}-\frac 12 h_{kk,ij}-\frac
12\frac 1{\vec \partial ^2}h_{kl,klij}). \label{TT}
\end{eqnarray}
It can be easily checked that both $h^{TT}_{ij}$ and
$h^T_{ij}$ are invariant under the linear
gauge-transformation $\delta h_{ij}=\xi_{i,j}+\xi_{j,i}$
\cite{MTW}.

Schwinger also formulated a TT decomposition, by a slightly
different parametrization \cite{Schw63}:
\begin{equation}
h_{ij}=h_{ij}^{TT} +\frac 12 (q_{i,j}+q_{j,i})-\delta_{ij}
q_{k,k}+q_{,ij}. \label{SchwTT}
\end{equation}

Here the terms containing $q_i$ are constructed to be
``doubly transverse'': $[\frac 12
(q_{i,j}+q_{j,i})-\delta_{ij} q_{k,k}]_{,ij}=0$. As a
result, $q$ is given by $(\vec \partial^2)^2 q=h_{ij,ij}$.
Then by examining the trace and divergence of Eq.
(\ref{SchwTT}) one obtains
\begin{equation}
q_i=\frac 1{\vec \partial^2} (2h_{ij,j}-\frac 12
h_{jj,i}-\frac 32 \frac 1{\vec \partial^2} h_{jk,jki} ).
\label{qi}
\end{equation}

The TT component $h_{ij}^{TT}$ is the same as that in Eq.
(\ref{TT}).

Later, Deser presented a covariant decomposition
\cite{Dese67}:
\begin{equation}
h_{ij} =\psi_{ij}^T +\nabla_i V_j +\nabla_j V_i.
\label{Deser}
\end{equation}
Here $\psi_{ij}^T$ is covariantly transverse: $\nabla^i \psi_{ij}^
T=0$. Deser does not extract further a TT part since the covariant
extension of the linear $h^T_{ij}$ is rather involving. The unknown
covariant vector $V_i$ is to be solved iteratively, and the leading
term is just $f_i$ in (\ref{fi}).

By a clever parametrization, York obtained a covariant TT
decomposition \cite{York73}:
\begin{equation}
h_{ij} =h_{ij}^{ TT} +(\nabla_i W_j +\nabla_j W_i-\frac 23
g_{ij} \nabla_k W^k )-\frac 13 g_{ij} h^k_{~k}.
\end{equation}
Here the middle term is constructed to be traceless. Again,
the unknown covariant vector $W_i$ is to be solved
iteratively. The leading term is
\begin{equation}
W_i =\frac 1{\vec \partial ^2}(h_{ij,j} -\frac 14 h_{jj,i}
-\frac 14 \frac 1{\vec \partial ^2} h_{jk,jki}), \label{Wi}
\end{equation}
and the leading TT component is the same as that in Eq.
(\ref{TT}).

The CZ decomposition has a very different formulation
\cite{Chen11d}. It applies to space-time instead of just
space, and is designed to be a clean separation of
gravitational and pure-gauge degrees of freedom up to
moderately strong field. The defining equations are
\begin{subequations}
\label{CZ0}
\begin{eqnarray}
g_{\mu\nu}\equiv \hat g_{\mu\nu} +\bar g_{\mu\nu};\\
\bar R^\rho_{~\sigma \mu\nu} (\bar g_{\alpha\beta})=0, \\
g^{ij} \hat \Gamma^\rho_{ij}=0.
\end{eqnarray}
\end{subequations}
Here $\bar g_{\mu\nu}$ is intended as a pure-gauge
background metric. Namely, it has an inverse $\bar
g^{\mu\nu}$ and can be used to define a background
connection $\bar \Gamma ^\rho_{\mu\nu}\equiv \frac 12 \bar
g^{\rho\sigma} (\partial_\mu \bar g_{\sigma \nu}
+\partial_\nu \bar g_{\sigma\mu} -\partial_\sigma \bar
g_{\mu\nu})$ with a vanishing Riemann curvature tensor
$\bar R^\rho_{~\sigma \mu\nu}\equiv
\partial_\mu \bar\Gamma^\rho_{\sigma \nu}-\partial _\nu \bar\Gamma
^\rho_{\sigma \mu} +\bar\Gamma ^\rho_{\alpha \mu}\bar\Gamma
^\alpha _{\sigma \nu} -\bar\Gamma ^\rho_{\alpha \nu}
\bar\Gamma ^\alpha _{\sigma \mu}=0$. The intended physical
term $\hat g_{\mu\nu}$ satisfies a delicate constraint
$g^{ij}\hat \Gamma ^\rho_{ij}=0$. Here $\hat \Gamma
^\rho_{\mu\nu}=\Gamma ^\rho_{\mu\nu}-\bar \Gamma
^\rho_{\mu\nu}$ is not an affine connection, and $\hat
g^{\mu\nu}\equiv g^{\mu\nu} -\bar g^{\mu\nu}$ is not the
inverse of $\hat g_{\mu\nu}$.

The CZ decomposition is also to be solved iteratively. At
linear order, $g_{\mu\nu}\equiv \eta_{\mu\nu}+h_{\mu\nu}$
with $\eta_{\mu\nu}$ the Minkowski metric and
$|h_{\mu\nu}|\ll 1$, an elegant, gauge-invariant expression
is obtained after a fairly lengthy calculation
\cite{Chen11d}:
\begin{equation}
\hat h_{\mu\nu}^{\rm CZ}=h_{\mu\nu}-\frac 1{\vec
\partial ^2}(h_{\mu i,i\nu} +h_{\nu
i,i\mu}-h_{ii,\mu\nu}).\label{CZ}
\end{equation}
Its trace is of special use and worth recording:
\begin{eqnarray}
\hat h_{ii}^{\rm CZ} =2(h_{ii}-\frac 1{\vec \partial ^2}
h_{ij,ij})=2\frac 1{\vec \partial ^2}  \partial_j
(h_{ii,j}-h_{ji,i}). \label{hii}
\end{eqnarray}
Moreover, an elegant relation can be derived for $\hat
h_{ij}^{\rm CZ}$ and the (fairly complicated) $h_{ij}^{TT}$
\cite{Chen11d}:
\begin{equation}
h_{ij}^{TT}=\hat h^{\rm CZ}_{ij} - \frac 14 (\delta _{ij}
\hat h ^{\rm CZ}_{kk} +\frac 1{\vec \partial ^2} \hat h
^{\rm CZ}_{kk,ij}).
\end{equation}

\section{Derivation of the linear decomposition by gauge-fixing}

Before discussing the highly tricky difference between
Deser, York, and CZ at higher orders, we first take a
closer look at their linear-order results, which we
rederive from a convenient gauge-fixing approach. The line
of our derivation will show clearly the {\em one-to-one
correspondence between complete gauge conditions and
gauge-covariant decompositions.}

Consider the linear gauge-transformation: $h'_{\mu\nu}
=h_{\mu\nu} -\partial_\mu \xi_\nu -\partial_\nu \xi_\mu$.
We show that the gauge parameter $\xi_\mu$ that brings
$h'_{\mu\nu} (x)$ into the ``generalized'' transverse
gauge,
\begin{subequations}
\label{ab}
\begin{eqnarray}
\partial_i h'_{i 0} +a\partial_0 h'_{ii} =0, \label{0}\\
\partial_i h'_{ij} +b\partial_j h'_{ii} =0, \label{i}
\end{eqnarray}
\end{subequations}
is essentially unique. Here $a,b$ can take any value except
$b= -1$, as Eq. (\ref{hii}) indicates that $(h_{j
i,i}-h_{ii,j})$ has a gauge-invariant divergence, thus
cannot be used to define a gauge. Various combinations of
$a,b$ have been studied in the literature. The gauge
$a=-\frac 14, b=0$ was employed by ADM \cite{ADM}, who also
suggested an important choice of $a=-\frac 12, b=-\frac
13$, which agrees with the Dirac gauge at linear
approximation \cite{Dira59}. The gauge $a=-\frac 23,
b=-\frac 13$ was encountered by Weinberg and termed ``too
ugly to deserve a name'' \cite{Wein65}, while the special
properties of $a=b=-\frac 12$ was recently revealed by CZ
\cite{Chen11g}.

We cast Eqs. (\ref{ab}) into the equations for $\xi_\mu$:
\begin{subequations}
\label{xi}
\begin{eqnarray}
\vec \partial ^2 \xi_0 +(1+2a ) \partial_0 \partial_i \xi_i
=\partial_i h_{i0}+a \partial_0 h_{ii},
\label{xia}\\
\vec \partial ^2 \xi_j +(1+2b ) \partial_j \partial_i \xi_i
=\partial_i h_{ij}+b \partial_j h_{ii}.\label{xib}
\end{eqnarray}
\end{subequations}
To solve, act on both sides of (\ref{xib}) with
$\partial_j$ and sum over $j$, we get
\begin{equation}
\xi_{i,i} = \frac 1{2(1+b)}(b h_{ii}+ \frac 1 {\vec
\partial^2} h_{ij,ij}).
\end{equation}
We remind that inversion of the Laplacian operator $\vec
\partial ^2$ implies a vanishing boundary value at
infinity. Substituting $\xi_{i,i}$ back into Eqs.
({\ref{xi}), we get
\begin{subequations}
\label{Sxi}
\begin{eqnarray}
\xi_0 =\frac 1{\vec \partial ^2}[h_{0 i,i} +\frac
{2a-b}{2(1+b)} h_{ii,0} -\frac {1+2a}{2(1+b)} \frac 1{\vec
\partial ^2} h_{ik,ik0}], \label{Sxia}\\
\xi_j =\frac 1{\vec \partial ^2}[h_{ij,i} +\frac b{2(1+b)}
h_{ii,j} -\frac {1+2b}{2(1+b)} \frac 1{\vec
\partial ^2} h_{ik,ikj}].  \label{Sxib}
\end{eqnarray}
\end{subequations}
These are the unique $\xi_\mu$ that bring $h'_{\mu\nu}$
into the gauge (\ref{ab}). The above derivation also show
clearly that Eqs. (\ref{ab}) fix the gauge completely: If
$h_{\mu\nu}$ is already in this gauge, then we get $\xi_\mu
\equiv 0$. Namely, no gauge-transformation can preserve
Eqs. (\ref{ab}).

The uniqueness of $\xi_\mu$ suggests the definition of a
gauge-invariant tensor, $\hat h_{\mu\nu}^{(ab)} =h_{\mu\nu}
-\xi_{\mu,\nu}-\xi_{\nu,\mu}$, which satisfies $\hat h_{0
i,i}^{(ab)} +a \hat h_{ii,0}^{(b)} =0$ and $\hat h_{j
i,i}^{(b)} +b \hat h_{ii,j}^{(b)} =0$. (We put a
superscript $^{(ab)}$ to remind the dependence on the
parameters $a,b$. The purely spatial components actually
involve only $b$, and $^{(b)}$ is used for clarity.) The
explicit expression of $\hat h_{\mu\nu}^{(ab)}$ is
\begin{subequations}
\label{hath}
\begin{eqnarray}
\hat h_{00}^{(ab)} &=&h_{00} -\frac 1{\vec
\partial ^2}(2h_{k0,k0}-h_{kk,00})\nonumber \\
&&-\frac{1+2a}{1+b} \frac 1{\vec
\partial ^2}( h_{kk} -\frac 1{\vec
\partial ^2} h_{kl,kl})_{,00},\label{00}\\
\hat h_{0j}^{(ab)} &=&h_{0j} -\frac 1{\vec
\partial ^2}(h_{k0,kj}+ h_{kj,k0}-h_{kk,0j})\nonumber \\
&&-\frac{1+a+b}{1+b} \frac 1{\vec
\partial ^2}( h_{kk} -\frac 1{\vec
\partial ^2} h_{kl,kl})_{,0j} \label{0j},\\
\hat h_{ij}^{(b)} &=&h_{ij} -\frac 1{\vec
\partial ^2}(h_{ki,kj}+ h_{kj,ki}-h_{kk,ij})\nonumber \\
&&-\frac{1+2b}{1+b} \frac 1{\vec
\partial ^2}( h_{kk} -\frac 1{\vec
\partial ^2} h_{kl,kl})_{,ij}. \label{hij}
\end{eqnarray}
\end{subequations}
We have organized $\hat h_{\mu\nu}^{(ab)}$ in a form which
displays evident gauge-invariance. By Eqs. (\ref{hath}),
the tensor $h_{\mu\nu} $ is separated into a
gauge-invariant part, $\hat h_{\mu\nu }^{(ab)}$, plus a
pure-gauge part, $\bar h_{\mu\nu} =h_{\mu\nu }-\hat
h_{\mu\nu}^{(ab)}=\xi_{\mu,\nu} +\xi_{\nu,\mu}$. Notice
that this is an operational separation since $\hat
h_{\mu\nu}^{(ab)}$ and $\bar h_{\mu\nu}$ are both expressed
explicitly in terms of the given $h_{\mu\nu}$. We thus see
that each complete gauge condition
$h_{i0,i}+ah_{ii,0}=0,h_{ij,i}+bh_{ii,j}=0$ (specified by
$a,b$) corresponds to one ways of decomposing $h_{\mu\nu}$
into an invariant part plus a pure-gauge part.

The expressions for $\hat h_{\mu\nu}^{(ab)}$ look rather
complicated, but the spatial trace $\hat h_{ii}^{(b)}$ is
remarkably simple for all choice of $b$:
\begin{eqnarray}
\hat h_{ii}^{(b)} =\frac 1{1+b} (h_{ii}-\frac 1{\vec
\partial ^2} h_{ij,ij}) =\frac 1{2(1+b)} \hat h_{ii}^{\rm
CZ}.
\end{eqnarray}

The gauge-invariant $\hat h_{\mu\nu}^{(ab)}$ relates to the
gauge-invariant $\hat h_{\mu\nu}^{\rm CZ}$ by
\begin{subequations}
\label{ab-CZ}
\begin{eqnarray}
\hat h_{00}^{(ab)}&=&\hat h_{00}^{\rm CZ} -\frac
{1+2a}{2(1+b)}\frac 1 {\vec
\partial ^2} (\hat h_{kk}^{\rm CZ})_{,00},\\
\hat h_{0j}^{(ab)}&=&\hat h_{0j}^{\rm CZ} -\frac
{1+a+b}{2(1+b)}\frac 1 {\vec
\partial ^2} (\hat h_{kk}^{\rm CZ})_{,0j},\\
\hat h_{ij}^{(b)}&=&\hat h_{ij}^{\rm CZ} -\frac
{1+2b}{2(1+b)}\frac 1 {\vec
\partial ^2} (\hat h_{kk}^{\rm CZ})_{,ij}.
\end{eqnarray}
\end{subequations}
Another important relation is between $h^{TT}_{ij}$ and
$\hat h_{ij}^{(b)}$:
\begin{equation}
h^{TT}_{ij}=\hat h_{ij}^{(b)}-\frac {1+b}2\delta_{ij}
h^{(b)}_{kk} +\frac {1+3b}2 \frac 1{\vec\partial^2}
h^{(b)}_{kk,ij}. \label{ab-TT}
\end{equation}
This agrees with Eq. (\ref{TT}) in the gauge $h_{ij,i}+b
h_{ii,j}=0$.

For comparison, we recall the parallel construction for a
vector field $A^\mu$ with the gauge-transformation $A'_\mu
(x) =A_\mu (x) -\partial_\mu \Lambda (x)$. The parameter
$\Lambda$ that brings $A'_\mu$ into the Coulomb gauge,
$\partial_i A'_i=0$, is also unique: $ \Lambda=\frac 1
{\vec \partial ^2} A_{i,i} $. This says that Coulomb gauge
is a complete gauge for a vector field. We can again define
a gauge-invariant quantity, $\hat A_\mu =A_\mu
-\partial_\mu \frac 1 {\vec \partial ^2} A_{i,i}$; and make
the decomposition $A_\mu\equiv \hat A_\mu+\bar A_\mu$, with
$\bar A_\mu=\partial_\mu \frac 1 {\vec \partial ^2}
A_{i,i}$ being a pure-gauge. The spatial components, $\vec
{\hat A}$ and $\vec {\bar A}$, are nothing but the
transverse field $\vec A_\perp$ and the longitudinal field
$\vec A_\parallel$.

Eqs. (\ref{Sxi}) and ({\ref{hath}) are far more complicated
than their counterparts for vector field, not only by more
indices, but by a doubly-nonlocal operation $\frac 1{\vec
\partial ^2} \frac 1{\vec \partial ^2}$ which also appears
in (\ref{fi}), (\ref{TT}), (\ref{qi}) and (\ref{Wi}).

Some special choices of $a,b$ are particularly interesting:

(i) $a=b=-\frac 12$ kills the doubly-nonlocal terms in Eqs.
(\ref{Sxi}) and ({\ref{hath}), and gives the CZ
decomposition in Eq. (\ref{CZ}), with the pure-gauge part:
\begin{subequations}
\begin{eqnarray}
\bar h_{\mu\nu}^{\rm CZ} &=& \frac 1{\vec
\partial
^2}(h_{\mu i,i\nu} +h_{\nu i,i\mu}-h_{ii,\mu\nu})\nonumber
\\ &\equiv& \xi^{\rm
CZ}_{\mu,\nu}+\xi^{\rm CZ}_{\nu,\mu},\\
\xi^{\rm CZ}_\mu &=&\frac 1{\vec \partial ^2}(h_{\mu i,i}
-\frac 12 h_{ii,\mu}).
\end{eqnarray}
\end{subequations}

(ii) $b=0$ gives a gauge-transformation parameter $\xi_i$
equal to $f_i$ in Eq. ({\ref{fi}), and thus the pure-gauge
part of the ADM decomposition of a spatial tensor:
\begin{subequations}
\begin{eqnarray}
\bar h_{ij}^{\rm ADM} &=& \frac 1{\vec \partial
^2}(h_{ik,kj} +h_{jk,ki}-\frac 1{\vec \partial
^2}h_{kl,klij})\nonumber\\
&\equiv & \xi^{\rm ADM}_{i,j}+\xi^{\rm ADM}_{j,i},\\
\xi^{\rm ADM}_i &=&f_i=\frac 1{\vec \partial ^2}(h_{i k,k}
-\frac 12 \frac 1{\vec \partial ^2} h_{kl,kli}).
\end{eqnarray}
\end{subequations}
This is also the linear-order result of
the covariant Deser decomposition in Eq. (\ref{Deser}). The
choice $b=0$ is not as convenient as one may expect, as the
doubly-nonlocal structure remains. This is similar to the
familiar fact in gravity that the gauge $\partial^\mu
h_{\mu\nu}=0$ is not convenient, and a cleverer choice is
the harmonic gauge $h^\mu_{~\nu,\mu}-\frac 12
h^\mu_{~\mu,\nu}=0$.

The two pure-gauge terms $\bar h_{ij}^{\rm ADM}$ and $\bar
h_{ij}^{\rm CZ}$ differ by another pure-gauge term:
\begin{equation}
\bar h_{ij}^{\rm ADM}=\bar h_{ij}^{\rm CZ}+\frac 12 \frac 1
{\vec
\partial ^2} (\hat h_{kk}^{\rm CZ})_{,ij}.
\label{CZ-ADM}
\end{equation}

(iii) $b=-\frac 13$ gives a gauge-transformation parameter
$\xi_i$ equal to $W_i$ in Eq. (\ref{Wi}), and thus the
pure-gauge part of York decomposition at linear order:
\begin{subequations}
\begin{eqnarray}
\bar h_{ij}^{\rm York}&=&\frac 1{\vec \partial ^2}(h_{i
k,kj} +h_{j k,ki}-\frac 12 h_{kk,ij} -\frac 12\frac 1{\vec
\partial ^2}h_{kl,klij}) \nonumber \\
&\equiv &\xi^{\rm York}_{i,j}+\xi^{\rm York}_{j,i}, \\
\xi^{\rm York}_i &=&W_i=\frac 1{\vec \partial ^2}(h_{i k,k}
-\frac 14 h_{kk,i} -\frac 14 \frac 1{\vec \partial ^2}
h_{kl,kli}).
\end{eqnarray}
\end{subequations}

$\bar h_{ij}^{\rm York}$ differs from $\bar h_{ij}^{\rm
CZ}$ by yet another pure-gauge term:
\begin{equation}
\bar h_{ij}^{\rm York}=\bar h_{ij}^{\rm CZ}+\frac 14 \frac
1 {\vec \partial ^2} (\hat h_{kk}^{\rm CZ})_{,ij}.
\label{CZ-York}
\end{equation}

When applied to general relativity, the pure-gauge term
$\bar h_{\mu\nu}=\xi_{\mu,\nu} +\xi_{\nu,\mu}$ relates to
the inertial effect associated with coordinate choice. They
{\em differ} in the decompositions of ADM (Deser), York,
and CZ. Accordingly, the gauge-invariant parts in these
decompositions, which are essentially the tensors defined
in the gauges $h_{j i,i}=0$, $h_{j i,i}-\frac 13
h_{ii,j}=0$, and $h_{j i,i}-\frac 12 h_{ii,j}=0$,
respectively, also differ. We will discuss further the
properties of these different gauge-invariant quantities in
later sections.

As far as gauge transformation is concerned, the CZ gauge
with $a=b=-\frac 12$ is the most convenient one: It relates
to all gauge configurations by just a singly-nonlocal
manipulation, while doubly-nonlocal manipulation may be
needed when transiting between two other gauges. The choice
$b=-\frac 13$, however, possesses a most important feature
that it kills the non-local term in Eq. (\ref{ab-TT}), thus
simplifies the extraction of $h_{ij}^{TT}$ to be just
algebraic. This feature leads to various important
applications:

(iv) Weinberg combines $b=-\frac 13$ with $a=-\frac 23$,
which simplifies $h_{ij}^{TT}$ and kills the
doubly-nonlocal term in Eq. (\ref{0j}). By this choice
Weinberg was able to demonstrate quantum Lorentz invariance
for the gravitational coupling \cite{Wein65}.

(v) The linearized Dirac gauge is to combine $b=-\frac 13$
with $a=-\frac 12$. It simplifies $h_{ij}^{TT}$ and kills
the doubly-nonlocal term of the time-time component in Eq.
(\ref{00}), which closely relates to the Newtonian
potential. This gauge is thus especially advantageous in
post-Newtonian dynamics \cite{Jaro98,Damo01}, and is also
referred to as the ADM-TT gauge. It should be clarified
that the term ``TT gauge" here just means that in this
gauge the TT component $h_{ij}^{TT}$ relates to $h_{ij}$
{\it locally}: $h_{ij}^{TT}=h_{ij}-\frac 13 \delta_{ij}
h_{kk}$, not that $h_{ij}$ is purely TT.

\section{Beyond the linear order}
The difference between Deser, York, and CZ beyond the
linear order, on the other hand, is much more drastic. The
delicate formulation of CZ via the Riemann curvature
guarantees that $\bar g_{\mu\nu}$ is a pure-gauge
background to all orders. However, in the Deser or York
formulation, $\nabla_i V_j+\nabla_j V_i$ or $\nabla_i
W_j+\nabla_j W_i$ is not a pure-gauge beyond linear order,
namely, it does not give a vanishing Riemann curvature. A
correct parametrization of a pure-gauge background metric
can be obtained through coordination transformation
$x'^\mu=x^\mu+\xi^\mu$ from the Minkowski metric
$\eta_{\mu\nu}$:
\begin{eqnarray}
\bar g_{\mu\nu} &=&\frac{\partial x'^\rho}{\partial x^\mu}
\frac{\partial x'^\sigma }{\partial x^\nu} \eta_{\rho\sigma}\nonumber\\
&=&\eta_{\mu\nu} +\eta_{\rho\nu} \frac{\partial
\xi^\rho}{\partial x^\mu} +\eta_{\sigma \mu}\frac{\partial
\xi^\sigma}{\partial x^\nu} +\eta_{\rho\sigma}
\frac{\partial \xi^\rho}{\partial x^\mu}\frac{\partial
\xi^\sigma}{\partial x^\nu}. \label{explicit}
\end{eqnarray}

The Riemann curvature of this metric is identically zero:
$\bar R^\rho_{~\sigma\mu\nu}(\bar g_{\alpha\beta})\equiv
0$. In principle, one may use such a parametrization to
formulate a clean separation of a pure-gauge background,
equivalent to what CZ achieve:
\begin{subequations}
\begin{eqnarray}
g_{\mu\nu} \equiv \hat g_{\mu\nu}+\eta_{\mu\nu}
+\eta_{\rho\nu} \xi^\rho_{,\mu} +\eta_{\sigma\mu}
\xi^\sigma_{,\nu} +\eta_{\rho\sigma}
\xi^\rho_{,\mu}\xi^\sigma_{,\nu};\\
g^{ij}\{\Gamma ^\rho_{ij}-\bar \Gamma ^\rho_{ij}[\bar
g_{\alpha\beta}(\xi^\lambda)]\}=0.
\end{eqnarray}
\end{subequations}
It can be seen, however, that this formulation is no less
demanding than that of CZ, since (except at linear order)
the equations for $\xi^\mu$ are even more involving than
those for $\bar g_{\mu\nu}$.

Analogously, when decomposing a Yang-Mills field:
$A^\mu=\bar A^\mu +\hat A^\mu$, it is not really easier to
parameterize the pure-gauge field $\bar A^\mu$ {\em
explicitly} as $U\partial ^\mu U^{-1}$ and try to solve
$U=e^{i\omega ^a T^a}$, compared to defining $\bar A^\mu$
{\em implicitly} by a vanishing field strength $\bar
F^{\mu\nu}(\bar A^\rho)=0$ and solving $\bar A^\mu$ as Chen
{\it et al.} do \cite{Chen08,Chen09}. Though being tedious,
the field decomposition in Refs.
\cite{Chen08,Chen09,Chen11d} can indeed be solved
straightforwardly to any desired order.

We point out that the Deser decomposition can be modified
to pick out a clean pure-gauge background to all orders,
following the construction line of CZ:
\begin{subequations}
\label{Deser'}
\begin{eqnarray}
g_{\mu\nu} =\psi^T_{\mu\nu} +\bar g_{\mu\nu};\\
\bar R^\rho_{~\sigma\mu\nu}(\bar g_{\alpha\beta})=0,\\
\nabla^i \psi ^T_{i\mu}=0.
\end{eqnarray}
\end{subequations}
This would be equivalent to, but easier to solve than
parameterizing the background explicitly as in Eq.
(\ref{explicit}):
\begin{subequations}
\label{Deser''}
\begin{eqnarray}
g_{\mu\nu} \equiv \psi^T_{\mu\nu}+\eta_{\mu\nu}
+\eta_{\rho\nu} \xi^\rho_{,\mu} +\eta_{\sigma\mu}
\xi^\sigma_{,\nu} +\eta_{\rho\sigma}
\xi^\rho_{,\mu}\xi^\sigma_{,\nu};\\
\nabla^i \psi ^T_{i\mu}=0.
\end{eqnarray}
\end{subequations}

At linear order, Eqs. (\ref{Deser'}) or (\ref{Deser''})
give the same $\psi^T_{ij}$ as by Eq. (\ref{Deser}). But
beyond the linear order, Eqs. (\ref{Deser'}) or
(\ref{Deser''}) can guarantee that
$(g_{\mu\nu}-\psi^T_{\mu\nu})$ is a pure-gauge, while Eq.
(\ref{Deser}) cannot.

\section{Equations of motion for the gauge-invariant fields}

To look into the physical implications of the tensor
decompositions, we derive here the equations of motion for
the various gauge-invariant quantities constructed by ADM,
Deser, York, and CZ. As we will see, this provides a very
illuminating perspective on the dynamics of general
relativity, and sheds important light on the question of
what are the most appropriate physical variables (or
equivalently, the most appropriate gauge) for gravitational
field.

The unconstrained (gauge-dependent) $h_{\mu\nu}$ satisfies
the linearized Einstein equation
\begin{equation}
\square h_{\mu\nu}-\partial_\mu
\partial_\rho h^\rho_{~\nu}-\partial_\nu\partial_\rho
h^\rho_{~\mu}+\partial_\mu\partial_\nu h^\rho_{~\rho}=-
S_{\mu\nu} . \label{Eins}
\end{equation}
Here $\square \equiv \vec \partial ^2 -\partial_t^2$,
$S_{\mu\nu} \equiv T_{\mu\nu}-\frac 12 \eta_{\mu\nu}
T^\rho_{~\rho}$, and we put $16\pi G=1$. Of the
gauge-invariant quantities built out of $h_{\mu\nu}$,
$h^{(ab)}_{\mu\nu}$ stands a primary role. Other quantities
like $h^{CZ}_{\mu\nu}$ or $\psi^T_{\mu\nu}$ are just
special cases of $h^{(ab)}_{\mu\nu}$ for certain $a,b$. We
first display the relevant equations for
$h^{(ab)}_{\mu\nu}$, then explain their derivations and
physical meanings:
\begin{subequations}
\label{Eqs-ab}
\begin{eqnarray}
\vec \partial ^2 \hat h_{00}^{(ab)}&=&-S_{00}
+\frac{1+2a}{1+b}\frac 1{\vec\partial^2} T_{00,00}
\nonumber\\&=&\vec
\partial^2 h_{00}\left|_{(h_{i0,i}+ ah_{ii,0}=0,~h_{ij,i}+
bh_{ii,j}=0)}\right. ,
\label{Eqa}\\
\vec \partial ^2 \hat h_{0i}^{(ab)}&=&-S_{0i}
+\frac{1+a+b}{1+b}\frac 1{\vec\partial^2} T_{00,0i}
\nonumber\\&=&\vec
\partial^2 h_{0i}\left|_{(h_{i0,i}+ ah_{ii,0}=0,~h_{ij,i}+
bh_{ii,j}=0)}\right. ,
\label{Eqb}\\
\vec \partial ^2\hat h^{(ab)}_{ii} &=& -\frac 1{1+b}T_{00}
=\vec
\partial ^2 h_{ii}\left|_{(h_{ij,i}+b h_{ii,j}=0)}\right. ,
\label{Eqc}\\
\vec \partial ^2\hat h^{(b)}_{ji,i} &=& \frac
b{1+b}T_{00,j}=\vec \partial ^2 h_{ji,i}\left|_{(h_{ij,i}+b
h_{ii,j}=0)}\right. ,
\label{Eqd}\\
\square \hat h_{ij}^{(b)}&=&-\hat S^{(b)}_{ij}=\square
h_{ij}\left|_{(h_{ij,i}+b h_{ii,j}=0)}\right. . \label{Eqe}
\end{eqnarray}
\end{subequations}

The source term $\hat S_{ij}^{(b)}$ is built with $S_{ij}$
in the same way as $\hat h_{ij}^{(b)}$ is built with
$h_{ij}$ in Eq. (\ref{hij}):
\begin{eqnarray}
\hat S_{ij}^{(b)} &=&S_{ij} -\frac 1{\vec
\partial ^2}(S_{ki,kj}+ S_{kj,ki}-S_{kk,ij})\nonumber \\
&&-\frac{1+2b}{1+b} \frac 1{\vec
\partial ^2}( S_{kk} -\frac 1{\vec
\partial ^2} S_{kl,kl})_{,ij}. \label{Sij}
\end{eqnarray}
The reason is that the equation for the gauge-invariant
quantity $ h_{ij}^{(b)}$ must be gauge-invariant, therefore
can be derived in any convenient gauge. Eqs. (\ref{Eqe})
and (\ref{Sij}) are direct consequence of the equation in
the harmonic gauge, $\square h_{\mu\nu}=-S_{\mu\nu}$.

We have organized Eqs. (\ref{Eqs-ab}) to show an important
and delicate {\em dual} relation between gauge-invariant
equations and gauge-fixed equations: If there exists an $X$
gauge in which a gauge-invariant quantity
$Y_{\mu\nu}=h_{\mu\nu}$, then the {\em gauge-invariant}
equations for the {\em gauge-invariant} $Y_{\mu\nu}$ take
the same form as the {\em gauge-dependent} equations for
the {\em gauge-dependent} $h_{\mu\nu}$ in the specific $X$
gauge. This dual relation can be very handy in deriving
equations. E.g., to seek the equations of motion for
$h_{\mu\nu}$ in the gauge $h_{i0,i}+a h_{ii,0}=0,~
h_{ij,i}+b h_{ii,j}=0$, we can instead look at the
equations of motion for the gauge-invariant quantity $\hat
h_{\mu\nu}^{(ab)}$ in Eqs. (\ref{hath}), which then can be
derived in the convenient harmonic gauge. Even more
conveniently, we see in Eqs. (\ref{ab-CZ}) that the
expression of $\hat h^{(ab)}_{\mu\nu}$ via $\hat h^{\rm
CZ}_{\mu\nu}$ is fairly simple, so we can derive the
equations of motion for $\hat h^{(ab)}_{\mu\nu}$ via the
equations of motion for $\hat h^{\rm CZ}_{\mu\nu}$, which
in turn are just the equations of motion for $h_{\mu\nu}$
in the gauge $h_{i\mu,i}-\frac 12 h_{ii,\mu}=0$, as we
recently derive in Ref. \cite{Chen11g}:
\begin{subequations}
\label{Eqs-CZ}
\begin{eqnarray}
\vec \partial ^2 \hat h_{0\mu}^{\rm CZ}&=&-S_{0\mu}=\vec
\partial ^2 h_{0\mu}\left|_{(h_{i\mu,i}- \frac 12
h_{ii,\mu}=0)}\right. ,
\\
\vec \partial ^2\hat h^{\rm CZ}_{ii} &=& -2T_{00} =\vec
\partial ^2 h_{ii} \left|_{(h_{i\mu,i}- \frac 12
h_{ii,\mu}=0)}\right. ,
\\
\vec \partial ^2\hat h^{\rm CZ}_{ji,i} &=& -T_{00,j} =\vec
\partial ^2 h_{ji,i}\left|_{(h_{i\mu,i}- \frac 12
h_{ii,\mu}=0)}\right.,\\
\square \hat h _{ij}^{\rm CZ}&=&-\hat S_{ij}^{\rm CZ}
=\square h_{ij}\left|_{(h_{i\mu,i}- \frac 12
h_{ii,\mu}=0)}\right. .
\end{eqnarray}
\end{subequations}

Like $\hat S_{ij}^{(b)}$, $\hat S_{ij}^{\rm CZ}$ is built
with $S_{ij}$ in the same way as $\hat h_{ij}^{\rm CZ}$ is
built with $h_{ij}$; and $\hat S_{ij}^{(b)}$ relates to
$\hat S_{ij}^{\rm CZ}$ in the same way as $h_{ij}^{(b)}$
relates to $h_{ij}^{\rm CZ}$:
\begin{subequations}
\label{Source}
\begin{eqnarray}
\hat S_{ij}^{\rm CZ}&=&S_{ij}-\frac 1{\vec\partial^2}
(S_{ik,kj}+S_{jk,ki}-S_{kk,ij}), \label{SCZ}\\
\hat S_{ij}^{(b)} &=&\hat S_{ij}^{\rm CZ}-\frac
{1+2b}{2(1+b)}\frac 1 {\vec
\partial ^2} (\hat S_{kk}^{\rm CZ})_{,ij} .\label{Sb}
\end{eqnarray}
\end{subequations}

For a consistency check: as $a=b=-\frac 12$, Eqs.
(\ref{Eqs-ab}) reduce to Eqs. (\ref{Eqs-CZ}), which are
just the special case of  Eqs. (\ref{Eqs-ab}) with the
simplest sources terms.

The instantaneous Laplacian operator $\vec\partial^2$ in
Eqs. (\ref{Eqa}-\ref{Eqd}) means that the gauge-invariant
quantities $\hat h_{0\mu}^{(ab)}$, $\hat h_{ii}^{(b)}$, and
$h^{(b)}_{ij,j}$ are {\em non-dynamical or
non-propagating}. Equivalently, in the gauge $h_{i0,i}+a
h_{ii,0}=0,~ h_{ij,i}+b h_{ii,j}=0$, the component
$h_{0\mu}$, the spatial trace $h_{ii}$ and the spatial
divergence $h_{ij,j}$, are non-dynamical. We give a note on
the equations for the spatial trace and divergence. They
are certainly consistent with, and can be carefully
reorganized from, the apparently ``propagating-looking''
equations in Eqs. (\ref{Eqe}). But a more elucidating way
to reveal the non-dynamical character of $h_{ii}$ and
$h_{ij,j}$ is through Eq. (\ref{Eqb}) and the constraint
$h_{i0,i}+a h_{ii,0}=0,~ h_{ij,i}+b h_{ii,j}=0$: First,
$h_{0 i,i}+a h_{ii,0}=0$ says that the trace $h_{ii}$ has
the same property as $h_{0 i}$, which is non-dynamical by
Eq. (\ref{Eqb}). Then, $h_{j i,i}+b h_{ii,j}=0$ says that
the spatial divergence $h_{ji,i}$ is non-dynamical as well.

Since $\hat h_{ii}^{(b)}$ and $h^{(b)}_{ij,j}$ are
non-dynamical, the truly dynamical component of $\hat
h_{ij}^{(b)}$ is its TT part. The TT part of $\hat
h_{ij}^{(b)}$ actually equals $h^{TT}_{ij}$ in Eq.
(\ref{TT}), because $\hat h_{ij}^{(b)}$ and $h_{ij}$ differ
by a pure-gauge term which does not contribute TT
component. The equation of motion for $h^{TT}_{ij}$ is thus
of special importance:
\begin{equation}
\square h^{TT}_{ij}=-S^{TT}_{ij}.
\end{equation}
The source term $S^{TT}_{ij}$ is the TT part of $S_{ij}$.
It relates to $S_{ij}$ and $S_{ij}^{(b)}$ in the same way
as $h_{ij}^{TT}$ relates to $h_{ij}$ and $h_{ij}^{(b)}$:
\begin{eqnarray}
S_{ij}^{TT}&=&S_{ij}-\frac 12 \delta _{ij}(S_{kk}-\frac
1{\vec \partial ^2} S_{kl,kl}) \nonumber \\&&-\frac 1{\vec
\partial ^2}(S_{ik,kj}+S_{jk,ki}-\frac 12 S_{kk,ij}-\frac
12\frac 1{\vec \partial ^2}S_{kl,klij}) \nonumber \\
&=&\hat S_{ij}^{(b)}-\frac {1+b}2\delta_{ij} S^{(b)}_{kk}
+\frac {1+3b}2 \frac 1{\vec\partial^2} S^{(b)}_{kk,ij}.
\label{STT}
\end{eqnarray}

It should be noted that the equation for the TT component
$h^{TT}_{ij}$ does not have a gauge-fixing dual. The reason
is that $h_{ij}$ cannot in general be reduced to contain
only TT component, except for a pure wave without source
\cite{Chen11g}. In this regard, the tensor gauge field
behaves rather peculiar that there are infinitely many
gauges which can remove all nonphysical (gauge) degrees of
freedom, but there is no gauge which can directly pick out
the dynamical (TT) component. (As we explained in Section
III, one can at best simply the extraction of $h_{ij}^{TT}$
to be algebraic and local.) Here one should notice the
difference between ``nonphysical'' and ``non-dynamical''.
E.g., in electrodynamics, the instantaneous Coulomb
potential is non-dynamical but physical, and must be
included into the total Hamiltonian of the system.
Similarly, for gravity the instantaneous Newtonian
interaction (contributed by $h_{0\mu}$ and $h_{ii}$) is
non-dynamical but physical, and must be included into the
total Hamiltonian, especially in a quantum theory
\cite{Wein65}.

In electrodynamics, the Coulomb gauge is the unique
constraint to pick out the two physical (and at the same
time dynamical) components $\vec A_\perp$. The above
peculiar feature of tensor gauge field leads to an
embarrassing fact that among the infinitely many complete
gauge conditions, $h_{i0,i}+a h_{ii,0}=0,~ h_{ij,i}+b
h_{ii,j}=0$, no single choice is superior in all aspects.
Equations (\ref{Eqs-ab}) indicate that by $a=b=-\frac 12$
we obtain the simplest form of equations for all components
of $h_{\mu\nu}$. However, we still have to extract the
dynamical component $h^{TT}_{ij}$, and Eq. (\ref{ab-TT})
indicates that $b=-\frac 13$ leads to the simplest
expression for $h^{TT}_{ij}$. In Ref. \cite{Wein65},
Weinberg chooses $b=-\frac 13$ together with $a=-\frac 23$,
which simplifies the equation for $h_{0j}$, but leaves a
complicated equation for $h_{00}$. Since $h_{00}$ is
closely related to the Newtonian potential, post-Newtonian
dynamics favors $a=-\frac 12$ which leads to the simplest
equation for $h_{00}$; but then one must choose between the
convenience in dealing with $h_{0j}$ (which favors
$b=-\frac 12$) or the dynamical component $h_{ij}^{TT}$
(which favors $b=-\frac 13$).

\section{another perspective on gauge condition and
gauge-field decomposition}

To understand better the trickiness in a tensor gauge
theory, in this section we approach the problem of
gauge condition and gauge-field decomposition from another
perspective. In the decomposition of gauge fields into an
invariant part plus a pure-gauge part,
\begin{equation}
A_i \equiv \hat A_i +\bar A_i , ~~~ h_{ij} \equiv  \hat
h_{ij} +\bar h_{ij}, \label{decomp}
\end{equation}
we had essentially followed in Sec. III a line of
constructing the pure-gauge fields $\bar A_i$ and $\bar
h_{ij}$. We now proceed by asking: What are the possible
means of constructing the invariant quantities $\hat A_i$
and $\hat h_{ij}$? For the vector field, we know that the
fundamental gauge-invariant {\it local} quantity is the
field strength $F_{ij}(A_i)=A_{j,i} -A_{i,j}$. (For
simplicity, we consider the linear theory, and first look
at the Euclidean space with positive-definite metric. The
extension to Minkowski space-time with indefinite metric
requires just a little bit caution.) The gauge-invariant
vector $\hat A_i$ must be built with $F_{ij}$, and the only
possible structure with the same dimension as $A_i$ is
$\frac 1{\vec\partial^2} F_{ji,j}$. For the tensor field,
the fundamental gauge-invariant local quantity is the
linearized Riemann curvature, $R_{ijkl}(h_{ij})=
h_{ik,jl}-h_{il,jk}-h_{jk,il}+h_{jl,ik}$, with which one
can build the gauge-invariant tensor $\hat h_{ij}$. One
finds now {\it two} possible structures with the same
dimension as $h_{ij}$: $\frac 1{\vec\partial^2} R_{ikjk}$
and $\frac 1{\vec\partial^2} \frac 1{\vec\partial^2}
R_{ikjl,kl}$.

To assist the analysis of decompositions in Eq.
(\ref{decomp}), we introduce the terms {\it competent
invariant} and {\it competent pure-gauge}. A competent
invariant is the gauge-invariant part of a gauge field,
which gives the same field strength (or curvature) as does
the full gauge field; and a competent pure-gauge should
give vanishing field strength (or curvature) and contain
all the gauge degrees of freedom (namely, it transforms in
the same as does the full gauge field). For a gauge field,
the rest part of a competent invariant must be a competent
pure-gauge, and the rest part of a competent pure-gauge
must be a competent invariant. In Eq. (\ref{decomp}), since
$\bar A_i$ and $\bar h_{ij}$ are required to be pure-gauge,
the gauge-invariant $\hat A_i$ and $\hat h_{ij}$ must
contribute the whole field strength or curvature, and thus
are necessarily competent invariants.

We now show that the gauge-invariant quantities, $\frac
1{\vec\partial^2} F_{ji,j}$, $\frac 1{\vec\partial^2}
R_{ikjk}$, and $\frac 1{\vec\partial^2} \frac
1{\vec\partial^2} R_{ikjl,kl}$, are all competent
invariants. Using the Bianchi identities,
\begin{subequations}
\begin{eqnarray}
F_{ij,k}+F_{jk,i}+F_{ki,j}=0,\\
R_{ijkl,m}+R_{ijlm,k}+R_{ijmk,l}=0,
\end{eqnarray}
\end{subequations} and using the symmetry properties of
$F_{ij}$ and $R_{ijkl}$, a slight algebra can prove that
$\frac 1{\vec\partial^2} F_{ji,j}$ gives the same field
strength as that of $A_i$, and that both $\frac
1{\vec\partial^2} R_{ikjk}$ and $\frac 1{\vec\partial^2}
\frac 1{\vec\partial^2} R_{ikjl,kl}$ give the same
curvature as that of $h_{ij}$:
\begin{subequations}
\begin{eqnarray}
F_{ij}(\frac 1{\vec\partial^2} F_{ji,j})=(\frac
1{\vec\partial^2} F_{kj,k})_{,i}-(\frac 1{\vec\partial^2}
F_{ki,k})_{,j}\nonumber
\\=\frac 1{\vec\partial^2}(F_{kj,i}+F_{ik,j})_{,k}=\frac
1{\vec\partial^2}(F_{ij,k})_{,k}=F_{ij}; \\
R_{ijkl}(\frac 1{\vec\partial^2} R_{ikjk}) = (\frac
1{\vec\partial^2} R_{imkm})_{,jl}-
(\frac 1{\vec\partial^2} R_{imlm})_{,jk} \nonumber\\
-(\frac 1{\vec\partial^2} R_{jmkm})_{,il}+ (\frac
1{\vec\partial^2} R_{jmlm})_{,ik} =R_{ijkl}; \\
R_{ijkl}(\frac 1{\vec\partial^2}\frac 1{\vec\partial^2}
R_{ikjl,kl})
=(\frac 1{\vec\partial^2}\frac 1{\vec\partial^2}R_{ipkq,pq})_{,jl} \nonumber\\
-(\frac 1{\vec\partial^2}\frac
1{\vec\partial^2}R_{iplq,pq})_{,jk}
-(\frac 1{\vec\partial^2}\frac 1{\vec\partial^2}R_{jpkq,pq})_{,il}\nonumber\\
+(\frac 1{\vec\partial^2}\frac
1{\vec\partial^2}R_{jplq,pq})_{,ik} =R_{ijkl};
\end{eqnarray}
\end{subequations}
where we have displayed some detail in derivation for
the vector case, and omitted similar steps for the tensor case.

Being competent invariants, $\frac 1{\vec\partial^2}
F_{ji,j}$ qualifies as $\hat A_i$ in Eq. (\ref{decomp}),
and both $\frac 1{\vec\partial^2} R_{ikjk}$ and $\frac
1{\vec\partial^2} \frac 1{\vec\partial^2} R_{ikjl,kl}$
qualify as $\hat h_{ij}$ in Eq. (\ref{decomp}). In fact,
from the explicit expressions,
\begin{subequations}
\begin{eqnarray}
\frac 1{\vec\partial ^2}F_{ji,j}=A_i-\frac 1{\vec\partial
^2} A_{j,ji},\\
\frac 1{\vec\partial^2} R_{ikjk} =h_{ij}-\frac
1{\vec\partial
^2}(h_{ik,kj}+h_{jk.ki}-h_{kk,ij}) ,\\
\frac 1{\vec\partial^2} \frac 1{\vec\partial^2}
R_{ikjl,kl}=h_{ij} -\frac 1{\vec\partial ^2}(
h_{ik,kj}+h_{jk.ki} \nonumber \\
-\frac 1{\vec\partial^2} h_{kl,klij}),
\end{eqnarray}
\end{subequations}
we see that $\frac 1{\vec\partial ^2}F_{ji,j}$ is just
$A_i^\perp$, $\frac 1{\vec\partial^2} R_{ikjk}$ is just
$\hat h_{ij}^{CZ}$, and $\frac 1{\vec\partial^2} \frac
1{\vec\partial^2} R_{ikjl,kl}$ is just $\hat
h_{ij}^{(b=0)}$.

One can now understand, in two ways, why a tensor gauge
field cannot be uniquely decomposed into an invariant part
plus a pure-gauge:

(i) Since $\frac 1{\vec\partial^2} R_{ikjk}$ and $\frac
1{\vec\partial^2} \frac 1{\vec\partial^2} R_{ikjl,kl}$ are
both competent invariants, so is the weighted combination,
$\alpha \frac 1{\vec\partial^2} R_{ikjk}+(1-\alpha) \frac
1{\vec\partial^2} \frac 1{\vec\partial^2} R_{ikjl,kl}$,
with $\alpha$ an arbitrary parameter.

(ii) The difference of two competent invariants makes a
peculiar {\it gauge-invariant pure-gauge}:
\begin{eqnarray}
\frac 1{\vec\partial^2} R_{ikjk}- \frac 1{\vec\partial^2}
\frac 1{\vec\partial^2} R_{ikjl,kl}=\frac
1{\vec\partial^2}(h_{kk}-\frac 1{\vec\partial^2}
h_{kl,kl})_{,ij} \nonumber \\\equiv
(\Xi_{,i})_{,j}+(\Xi_{,j})_{,i}, ~{\rm with}~\Xi=\frac 12
\frac 1{\vec\partial^2}(h_{kk}-\frac 1{\vec\partial^2}
h_{kl,kl}) \label{GIPG}
\end{eqnarray}
(where the latter expression is for later use). This
peculiar quantity (multiplied by any factor) can be shifted
arbitrarily between $\hat h_{ij}$ and $\bar h_{ij}$, while
preserving $\hat h_{ij}$ a competent invariant and $\bar
h_{ij}$ a competent pure-gauge. Eq. (\ref{hij}) exhibits
exactly such a feature.

The existence of a gauge-invariant pure-gauge also explains
the non-uniqueness of {\it complete} tensor gauge
condition. Since a competent pure-gauge carries all the
gauge degrees of freedom, a constraint would qualify as a
complete gauge condition were it able to set a competent
pure-gauge to be identically zero. For a vector gauge
field, the competent pure-gauge is unique: $\bar
A_i=A_i^\parallel=(\frac 1 {\partial^2} A_{j,j})_{,i}$.
This can be set identically zero by requiring $A_{j,j}=0$,
which is the unique constraint that can fix the (vector)
gauge completely. For a tensor gauge field, however, the
existence of infinitely many competent pure-gauges (with
the freedom of adding a gauge-invariant pure-gauge in Eq.
(\ref{GIPG}) multiplied by an arbitrary factor) leads to
infinitely many complete tensor gauge conditions. If
employing our earlier derivations, the competent pure-gauge
can be put in the form $\bar h_{ij}=\xi_{i,j}+\xi_{j,i}$,
where $\xi_j$ takes the same expression as in Eq.
(\ref{Sxib}) with a free parameter $b$. $\xi_j$ (and thus
$\bar h_{ij}$) can be made identically zero by the
constraint $h_{ij,j}+b h_{jj,i}=0$, which then qualifies as
a complete tensor gauge condition (in Euclidean space).

In Minkowski space-time, one needs to pay attention that
only the Laplacian operator has a well-defined inverse,
while the d'Alembert operator $\square = \vec
\partial ^2 -\partial_t ^2$ does not. (More exactly,
inverting the Laplacian operator requires just boundary
conditions, while inverting the d'Alembert operator would
also require initial conditions which are unavailable.)
Therefore, $\hat A_\mu$ is to be constructed with $\frac
1{\vec \partial^2} F_{j\mu,j}$, instead of $\frac
1{\square} F_{\nu\mu,\nu}$. Analogously, $\hat h_{\mu \nu}$
is to be constructed with $\frac 1{\vec\partial^2} R_{\mu k
\nu k}$ and $\frac 1{\vec\partial^2} \frac
1{\vec\partial^2} R_{\mu k \nu l,kl}$, instead of $\frac
1{\square} R_{\mu \rho \nu \rho}$ and $\frac 1{\square}
\frac 1{\square} R_{\mu \rho \nu \sigma ,\rho \sigma}$.
Another fact worth noting is that $\hat h_{\mu\nu}$ can
posses two free parameters, not just one. The reason is
that $\hat h_{\mu\nu}$ can be added by a gauge-invariant
pure-gauge $\bar h_{\mu\nu}\equiv
\xi_{\mu,\nu}+\xi_{\mu,\nu}$, where $\xi_0=\alpha \Xi_{,0}$
and $\xi_j=\beta \Xi_{,j}$, with $\Xi$ in Eq. (\ref{GIPG})
and $\alpha,\beta$ being two free parameters. Accordingly,
the complete tensor gauge conditions in Minkowski
space-time also contain two free parameters, as formulated
equivalently by $a$ and $b$ in Eqs. (\ref{ab}).

\section{Summary and Discussion}

In this paper we carefully examined and demonstrated why
decomposing a tensor is much more tricky than decomposing a
vector, even for the linear case. Concerning mathematical
structures, it is the CZ decomposition that has exact
correspondence to the simple (and unique) vector
decomposition. Namely, by a (singly) nonlocal construction,
a (uniquely) gauge-invariant field $\hat h_{\mu\nu}^{\rm
CZ}$ can be built out of $h_{\mu\nu}$. Like the transverse
vector current $\vec j_\perp=\vec j -\vec \partial \frac
1{\vec \partial^2}\vec \partial \cdot \vec j$, the source
for $\hat h_{\mu\nu}^{\rm CZ}$ also contains at most a
singly nonlocal structure.

If doubly nonlocal constructions are allowed, however, more
nontrivial possibilities emerge for a tensor (but not for a
vector). Especially, a peculiar quantity which is
gauge-invariant but at the same time a pure-gauge can be
constructed. It is essentially this peculiar quantity that
leads to infinitely many ways of fixing the gauge
completely, or equivalently, of decomposing the gauge field
into invariant and pure-gauge parts. (From our
demonstration, it is clear that such phenomena would arise
for all gauge fields of spin $\ge 2$.)

A further and vital complication for the tensor gauge field
is that its truly dynamical component (the TT part) does
not show up automatically after all gauge degrees of
freedom have been removed; and extraction of the TT part
favors a different gauge than that of CZ. Our observation
reveals vividly that tensor coupling is indeed
extraordinarily nontrivial, and requires further careful
studies, even at linear approximation.

Beyond the linear order, separation of a pure-gauge
background requires very careful formulation. Moreover, the
nonlinear Einstein equations are so hard to solve that, for
the purpose of achieving certain sort of convenience, one
may have to invent some particularly delicate field
variables and gauge conditions \cite{Kol09,Damo85,Ohta};
some gauges are even too complicated to put into explicit
forms \cite{Ohta}. It should be remarked, however, that the
issue of gauge choice may not be just a matter of
convenience. As Weinberg elaborated in Ref. \cite{Wein65},
{\it quantum} Lorentz invariance for gravitation might not
be guaranteed with an arbitrary gauge-fixing.

This work is supported by the China NSF Grants 10875082 and
11035003. X.S.C. is also supported by the NCET Program of
the China Ministry of Education.

\end{document}